\documentclass[preprint,aps,prl,showkeys,showpacs,onecolumn,preprintnumbers,10pt]{revtex4-1}
\usepackage{graphicx}
\usepackage{color}
\usepackage{amsmath}
\usepackage{subfig}
\usepackage{float}
\usepackage{multirow}
\usepackage{flafter}

\begin{document}

\title{Detailed analysis of the Bi-O pockets problem in $Bi_2Sr_2Ca_2Cu_3O_{10}$}
\author{J. A. Camargo-Mart\'inez}
\affiliation{Departamento de F\'isica, CINVESTAV-IPN, Av. IPN 2508, 07360 M\'exico}
\author{Diego Espitia}
\affiliation{Departamento de F\'isica, CINVESTAV-IPN, Av. IPN 2508, 07360 M\'exico}
\author{R. Baquero}
\affiliation{Departamento de F\'isica, CINVESTAV-IPN, Av. IPN 2508, 07360 M\'exico}

\begin{abstract}

The Bi-O pockets problem, namely, the appearance in theoretical ab initio calculations of the electronic band structure of Bi-cuprates of a pocket of states at the Fermi 
energy ($E_F$) that is attributed to states belonging to the Bi-O plane is an issue that still calls for more study. The Bi-O pockets are in contradiction with experiments. We have 
investigated the possible reasons for the disagreement. We checked that by using the experimental lattice and internal parameters without any optimization procedure, 
the Bi-O pockets do not appear at $E_F$ in agreement with experiment. Nevertheless, as pointed out by R. Kouba et al. [{\em Phys. Rev. B} {\bf 60}, 9321 (1999)] optimization is compulsory to a 
band structure calculation that will describe appropriately the electronic properties. But starting with the experimental parameters a further optimization procedure previous to 
the actual ab initio calculation leads to the Bi-O pockets. Doping with 25\% of Pb they disappear. From the several configurations that we have considered, we found two very simple 
ways in which the Bi-O pockets disappear without avoiding an optimization procedure previous to the calculation and without including a doping of any kind. In this paper, we report 
the effect of the slight displacement of the oxygen atom associated to the Sr-plane (O3) in the electronic properties of $Bi_2Sr_2Ca_2Cu_3O_{10}$ (Bi-2223) with tetragonal structure 
($I4/mmm$) using the Local Density Approximation (LDA). The slight displacement is performed after the system has been optimized. We determined the intervals of the O3 atomic 
positions for which calculations of the band structures show that the Bi-O bands emerge towards higher energies in agreement with the experimental results, thereby solving the Bi-O 
pockets problem. This procedure  does not introduce foreign doping elements into the calculation. Our calculations present a good agreement with the angle-resolved 
photoemission spectroscopy (ARPES) and nuclear magnetic resonance (NMR) experiments. The two options found differ in the character (metallic or nonmetallic) of the Bi-O plane. 
There are no experiments to the best of our knowledge that determine this character for Bi-2223.

\end{abstract}
\date{\today}
\pacs{ 74.72.Hs ; 71.20.-b  ; 71.18.+y ; 73.20.At}
\keywords{Bi-2223, Electronic structure, Band structure, Fermi surface. }


\maketitle

\section{Introduction}

The $Bi_2Sr_2Ca_2Cu_3O_{10}$ compound (Bi-2223) is a high-temperature superconductor (HTSC) which shows a transition to the superconducting state at $\sim110 K$~\cite{A1,A2}. There
are very few theoretical studies of this compound in the literature in spite of the fact that it is one of the most suitable HTSC materials for applications~\cite{A,B,C}. Also, 
experimental reports show that the composite $CdS/Bi$-$2223$ and the $Bi$-$2212/2223$ intergrowth single crystals display the reentrant superconducting behavior~\cite{O,ooo}.

In a previous work~\cite{3a} we calculated the electronic properties of Bi-2223 which show the Bi-O pockets problem meaning the presence of Bi-O bands 
at the Fermi energy ($E_F$) around the high symmetry point $\overline{M}$ in the irreducible Brillouin zone (IBZ). This problem appears in all the bismuth cuprates, 
a result which does not correspond to the experiment. This is an issue that has been present in the literature since long ago~\cite{3a,1a,1aA,2a,4a,5a,6a,6aA}. H. Lin et al. 
showed that this issue can be solved by doping the bismuth cuprates with Pb~\cite{7a,8a}. Also, V. Bellini et al., simulated a Bi-O plane terminated (001) surface in 
$Bi_2Sr_2Ca_1Cu_2O_{8}$ (Bi-2212) and obtained that the Bi-O pockets become less distinguishable~\cite{a9}.

As it is known atomic displacements in the crystalline structure generate considerable changes in the electronic properties. Herman et al. reported band structures for 
$Bi_2Sr_2Ca_1Cu_2O_{8}$ (Bi-2212) that were calculated by displacing the Bi and O atoms and found significant shifts in the bands~\cite{9a}. However, their band 
structures differ greatly from those obtained by other workers~\cite{2a}. 

Relying ourselves on this idea, in this paper we study the effect of different atomic displacements and found that the displacements of the oxygen atoms associated to the Sr plane represent 
a possible solution to the disagreement between theory and experiment on the Bi-O pockets problem.

\section{Method of Calculation}
In this paper the calculations were done using the full-potential linearized augmented plane wave method plus local orbital (FLAPW+lo)~\cite{10a} within the local density 
approximation (LDA) using the wien2k code~\cite{11a}. The core states are treated fully relativistically while for the valence states the scalar relativistic approximation is used. 
We used a plane-wave cutoff at $R_{mt}K_{max}$= 7.0 and for the wave function expansion inside the atomic spheres, a maximum value for the angular momentum of $l_{max}$= 12 with 
$G_{max}$= 25. We choose a $17\times17\times17$ k-space grid which contains 405 points within the IBZ. The muffin-tin sphere radii $R_{mt}$ 
(in atomic units) are chosen to be 2.3 for Bi, 2.0 for Sr, 1.9 for both Ca and Cu, and 1.5 for O (1.02 for the form II see below).

\section{Structures}

\begin{figure*}[hbt!]
\includegraphics[width=0.8\textwidth]{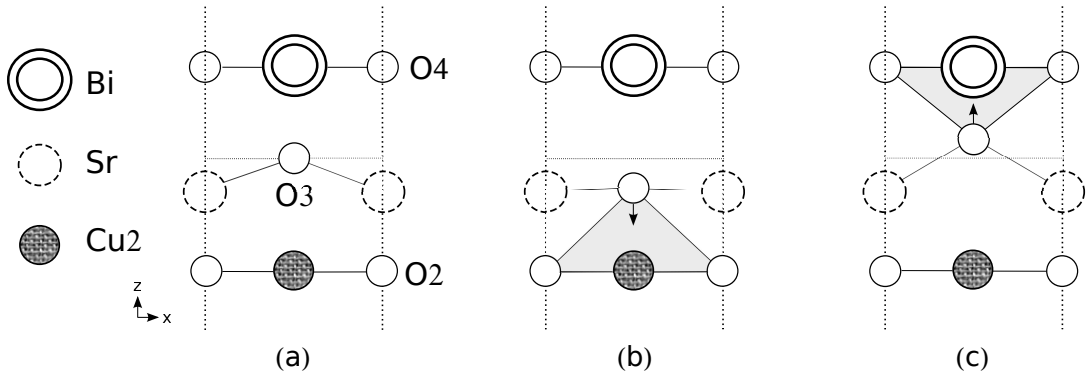} 
\caption{Schematic forms for the different atomic configurations a)  Form Opt, b) Form I and c) Form II.}
\label{formas}
\end{figure*}

In this work, we start from the Bi-2223 compound in a body center tetragonal structure (bct) and space group I4/mmm ($D_{4h}^{17}$). 
The structure consists of three Cu-O planes, one Cu1-O1 plane between two Cu2-O2 planes, with Ca atoms between them. Each Cu2-O2 plane is followed by 
a Sr-O3 and Bi-O4 planes in that order. In a previous work~\cite{3a} we optimized the $c/a$ ratio and relaxed the internal parameters of the 
structure taking as starting point the experimental internal parameters given in~\cite{12a}. Henceforth, we will refer to this optimized structure as form Opt 
(see Fig.~\ref{formas}(a)). Band structure calculations for the form Opt show the Bi-O pockets~\cite{3a}.

R. Kouba et al. found that one has to start from the theoretical equilibrium volume i.e. optimize the structure, because the use of the experimental lattice parameters represent an inconsistency~\cite{11bb}. 
We have checked that starting from the experimental lattice and internal parameters the Bi-O pockets do not appear at $E_F$ at the cost of avoiding the optimization procedure 
which as we just mentioned leads to a non optimal description of the electronic properties. 

Taking the optimized internal parameters of the structure as the starting point, we performed small displacements of the positions of the Bi, O4, O3 and Sr atoms in different 
configurations and found that some of them generate changes in the band structures that remove the Bi-O pockets. The simplest way is just to displace the O3 atoms in two 
small intervals around the optimized position (see Table~\ref{tablapos}). These intervals are defined in two structures which we labeled as form I and form II. It is important to note 
that O3 atomic positions out of these intervals show the Bi-O pockets back. We will discuss here the resulting electronic properties in two cases when the O3 atoms are displaced by
$\sim$0.68 $\AA$ towards the Cu2-O2 planes (form I) and when they are displaced by $\sim$0.27 $\AA$ towards the Bi-O4 planes (form II). See Fig.~\ref{formas}(b) and Fig.~\ref{formas}(c).

In Table~\ref{tablapos} we present the O3 atomic positions and the interatomic distances between Cu2 and O3 ($d_{Cu2-O3}$) and between Bi and O3 ($d_{Bi-O3}$) for the different configurations, and the intervals for the forms I and II. 
The atomic positions of the other atoms forming the crystal structure of Bi-2223 remain without changes and can be found in the ref.~\cite{3a}.

\begin{table}[hbt!]
\caption{\label{tablapos} The O3 atomic positions and the interatomic distances between Cu2 and O3 ($d_{Cu2-O3}$) and between Bi and O3 ($d_{Bi-O3}$) for the different configurations. The intervals for 
the forms I and II. 
The data for form Opt were taken from ref.~\cite{3a}. The experimental data were taken from ref.~\cite{12a}. }
\begin{tabular*}{0.4\textwidth}{@{\extracolsep{\fill}}lc|cc}\hline\hline
Structure&   O3 site     & \multicolumn{2}{c}{Interatomic distance ($\AA$)} \\\hline
         &    z/c        & $d_{Cu2-O3}$   & $d_{Bi-O3}$  \\
Exp     &    0.1454     &  1,7719   & 2,4281 \\
Form Opt &    0.1519     &  2.4946   & 2.0287 \\
Form I   &    0.1335     &  1.8196   & 2.7037 \\
Form II  &    0.1592     &  2,7624   & 1,7609 \\\hline
\multicolumn{4}{c}{Form I  Interval (z/c)     (0.133-0.134)}\\
\multicolumn{4}{c}{Form II Interval (z/c)     (0.1583-0.16)} \\
\hline\hline
\end{tabular*}
\end{table}

As it can be seen from Table~\ref{tablapos}, the displacements of O3 in the forms I and II produce important changes in the interatomic distances $d_{Cu2-O3}$ and $d_{Bi-O3}$. In the 
form I the $d_{Cu2-O3}$ is very close to the the experimental value (the difference is below 3.0\%), while in form II the $ d_ {Bi-O3} $ is $\sim0.67$ $\AA $ less than the 
experimental value. As we will see later these features will have consequences in the metallic or nonmetallic character of the Bi-O planes.

\section{Results and discussion}

\subsection{Band structure}

\begin{figure*}[htb!]
\includegraphics[width=0.7\textwidth]{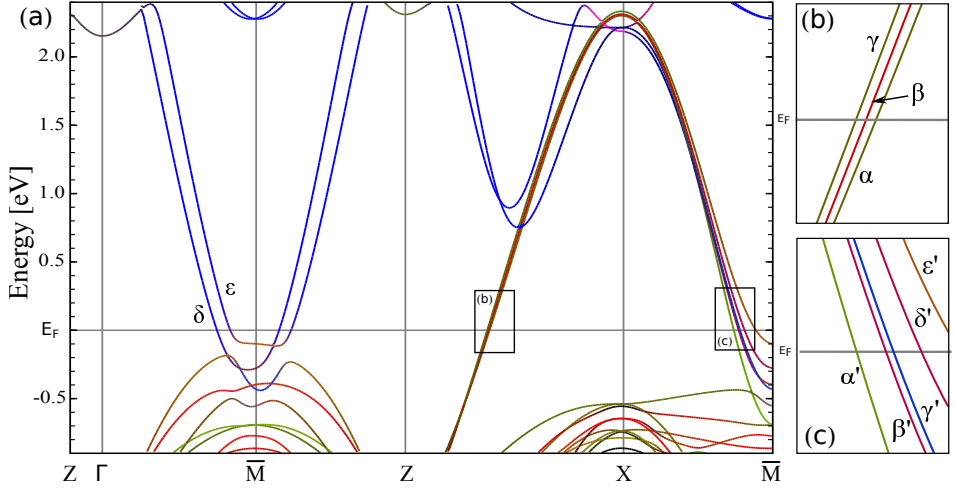}
\caption{(Color online) The band structure of the $Bi_2Sr_2Ca_2Cu_3O_{10}$ compound of the form Opt. The rectangles in the figure (a) are amplified in Figures (b) and (c) respectively. 
The blue, red and green lines represent the Bi $p$, Cu2 $d$, and Cu1 $d$ respectively. The hybridized states from these bands are represented by their respective color mixture 
and the black line represents the other states.}
\label{Fig.21} 
\end{figure*}

In this section we will present the band structures of the different crystalline configurations described in the previous section. The band structure of the form Opt is shown in 
Fig.\ref{Fig.21}. (Notice that $\overline{M}$ is the midpoint between the $\Gamma$ and Z along the $\Sigma$ direction). A detailed analysis of this band structure can be found in 
the ref.~\cite{3a}.

In the band structure of the form Opt we note the presence of five bands crossing at $E_F$ around the $\overline{M}$ point and in the Z-X direction (see Fig.\ref{Fig.21}(c)). These 
bands are derived from Cu1 $d$, Cu2 $d$, and Bi $p$ states. In the Z-X direction (see Fig.\ref{Fig.21}(b)) there are three bands labeled as $\alpha$, $\beta$, and $\gamma$, derived from 
the Cu-O planes, crossing at $E_F$ (which is characteristic of the HTSC cuprates). A relevant feature is the presence of Bi-O bands labeled as $\delta$ and $\varepsilon$ which drop 
below $E_F$ ($\sim 0.29 eV$) around the $\overline{M}$ point, hybridizing with the Cu-O bands. This behavior has never been observed experimentally~\cite{4a,5a,6a,6aA}.
The band dispersion in the $\Gamma$-Z direction (perpendicular to the basal plane) is minimal, which means that the bands are strongly two dimensional. 

\begin{figure*}[htb!]
\includegraphics[width=0.7\textwidth]{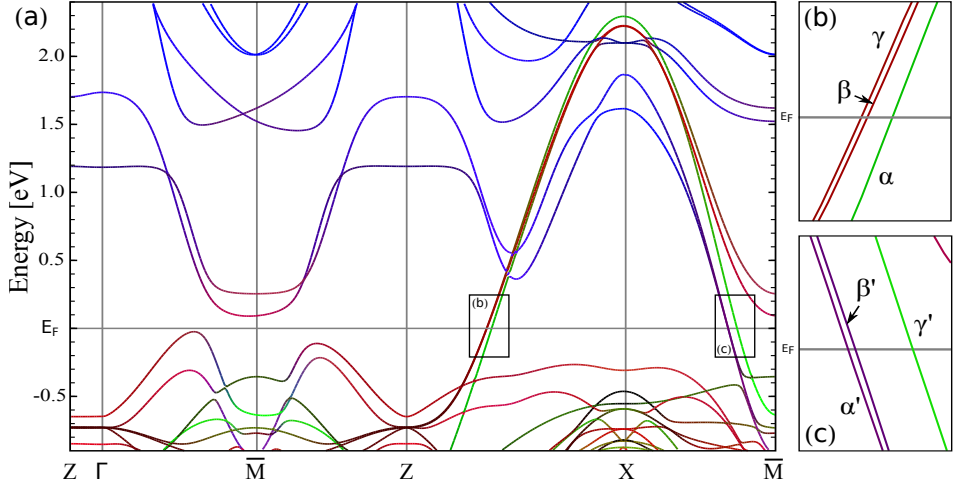}
\caption{(Color online) The band structure of the $Bi_2Sr_2Ca_2Cu_3O_{10}$ compound of the form I. The rectangles in the figure (a) are amplified in Figures (b) and (c) respectively. 
The blue, red and green lines represent the Bi $p$, Cu2 $d$, and Cu1 $d$ respectively. The hybridized states from these bands are represented by their respective color mixture 
and the black line represents the other states.}
\label{Fig.2} 
\end{figure*}

The band structure of the form I is shown in Fig.\ref{Fig.2}. We observed that the general behavior of this band structure of the form I is similar to the one calculated for the 
form Opt, although several important differences are present. It is observed in the band structure that only three bands cross at $E_F$ in two regions, in the Z-X direction and 
around the $\overline{M}$ point. In Table~\ref{Tabla2} we present in detail the contribution at $E_F$ from the different atomic states.

\begin{table*}[hbt!]
\caption{\label{Tabla2} Detailed contribution (form I) from the different atomic states to the bands at $E_F$.} 
\begin{tabular*}{0.95\textwidth}{@{\extracolsep{\fill}}cc | ccccccccccc}\hline\hline
                                              &                & \multicolumn{2}{c}{Bi} &    O4      & \multicolumn{2}{c}{O3} &     Cu1       & \multicolumn{2}{c}{O1} &     Cu2       &  \multicolumn{2}{c}{O2}  \\
                                                                \cline{3-4}        \cline{5-5}           \cline{6-7}       \cline{8-8}        \cline{9-10}         \cline{11-11}       \cline{12-13}
Direction                                     & Band           & $p_{x,y}$  & $p_{z}$   & $p_{x,y}$  & $p_{x,y}$  &  $p_{z}$  & $d_{x^2-y^2}$ &  $p_{x}$  &  $p_{y}$   & $d_{x^2-y^2}$ & $p_{x}$ &  $p_{y}$      \\\hline          
\multirow{3}{*}{Z-X}                          & $\alpha$       &     -      &    -      &      -     &      -     &     -     &     48\%      &    16\%   &    36\%    &     -         &   -     &    -          \\                       
                                              & $\beta$        &     -      &    -      &      -     &      11\%  &     -     &     -         &    -      &    -       &     33\%      &   20\%  &    36\%       \\
                                              & $\gamma$       &     -      &    -      &      -     &      11\%  &     -     &     -         &    -      &    -       &     33\%      &   20\%  &    36\%       \\\hline 
\multirow{3}{*}{X-$\overline{\text{M}}$} 
                                              & $\alpha^\prime$&     7\%    &    65\%   &      8\%   &      4\%   &     4\%   &     -         &    -      &    -       &     5\%       &   -     &    7\%       \\
                                              & $\beta^\prime$ &     18\%   &    8\%    &      21\%  &      10\%  &     10\%  &     -         &    -      &    -       &     15\%      &   -     &    18\%       \\
                                              & $\gamma^\prime$&     -      &    10\%   &     -      &      -     &     -     &     60\%      &    -      &    30\%    &     -         &   -     &    -         \\\hline                
\end{tabular*}
\end{table*}

In the first region (see Fig.\ref{Fig.2}(b)) there are three bands. Two of these bands are nearly degenerate (labeled as $\beta$ and $\gamma$); these are composed of Cu2 $d$, O2 
$p$, and O3 $p$ states. The other band (labeled as $\alpha$) is composed of Cu1 $d$ and, O1 (see Table~\ref{Tabla2}).

In the second region, in the X-$\overline{M}$ direction (see Fig.\ref{Fig.2}(c)) there are three bands crossing the $E_F$. Two of these bands are nearly degenerate (labeled as 
$\alpha^\prime$ and $\beta^\prime$), these are composed of Bi $p$, O4 $p$, O3 $p$, Cu2 $d$, and O2 $p$ states (see Table~\ref{Tabla2}). Here the metallic character of the Bi-O bands is evident.
The other band (labeled as $\gamma^\prime$) is composed of Cu1 $d$, O1 $p$ and Bi $p$ states.

The main difference is present around the $\overline{M}$ point in the $\Gamma$-Z direction (see Fig.\ref{Fig.2}(a)), where it is observed that the bands derived from hybrids Bi-O and
Cu-O states raise above $E_F$ $\sim 90 meV$, while the bands below $E_F$ (with the same composition that previous ones) reach $\sim -25 meV$.

It is observed in the band structure of the form I that some Bi-O bands do not follow a rigid displacement to the energies above $E_F$. Below $E_F$  important 
changes are observed in the dispersion of Cu-O bands with presence of Bi-O states around $\overline{M}$. 

\begin{figure*}[htb!]
\includegraphics[width=0.7\textwidth]{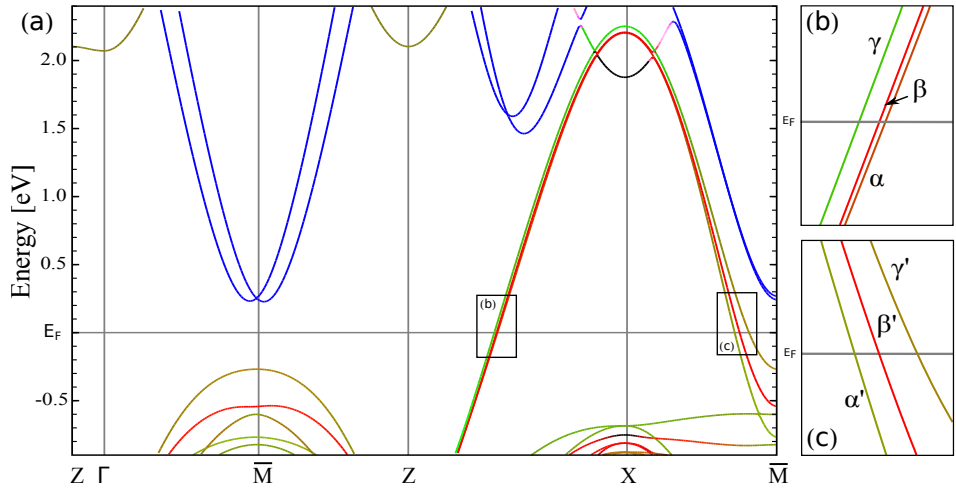}
\caption{(Color online) The band structure of the $Bi_2Sr_2Ca_2Cu_3O_{10}$ compound of the form II. The rectangles in the figure (a) are amplified in Figures (b) and (c) respectively. 
The blue, red and green lines represent the Bi $p$, Cu2 $d$, and Cu1 $d$ respectively. The hybridized states from these bands are represented by their respective color mixture 
and the black line represents the other states.}
\label{Fig.2a} 
\end{figure*}

The band structure of the form II is shown in Fig.\ref{Fig.2a}. Just as in the previous case the band structure of the form II is similar to the one calculated for the form Opt  
with some differences. Again it is observed in the band structure that three bands cross $E_F$ in two regions, in the Z-X direction and around the $\overline{M}$ point. In 
Table~\ref{Tabla3} we present in detail the contribution at $E_F$ from the different atomic states.

\begin{table*}[hbt!]
\caption{\label{Tabla3} Detailed contribution (form II) from the different atomic states to the bands at $E_F$.} 
\begin{tabular*}{0.95\textwidth}{@{\extracolsep{\fill}}cc | cccccc}\hline\hline
                                              &                &     Cu1       & \multicolumn{2}{c}{O1} &     Cu2       &  \multicolumn{2}{c}{O2}  \\
                                                                  \cline{3-3}        \cline{4-5}           \cline{6-6}           \cline{7-8}       
Direction                                     & Band           & $d_{x^2-y^2}$ &  $p_{x}$  &  $p_{y}$   & $d_{x^2-y^2}$ & $p_{x}$ &  $p_{y}$      \\\hline          
\multirow{3}{*}{Z-X}                          & $\alpha$       &     15\%      &    8\%    &    11\%    &     44\%      &   8\%   &    14\%          \\                       
                                              & $\beta$        &     -         &    -      &    -       &     69\%      &   11\%  &    20\%       \\
                                              & $\gamma$       &     40\%      &    10\%   &    22\%    &     19\%      &   -     &    9\%       \\\hline 
\multirow{3}{*}{X-$\overline{\text{M}}$} 
                                              & $\alpha^\prime$&     36\%      &    -      &    16\%    &     82\%      &   -     &    18\%       \\
                                              & $\beta^\prime$ &     -         &    -      &    -       &     15\%      &   -     &    18\%       \\
                                              & $\gamma^\prime$&     32\%      &    -      &    15\%    &     42\%      &   -     &   12\%       \\\hline                
\end{tabular*}
\end{table*}

In the first region (see Fig.\ref{Fig.2a}(b)) there are three bands. Two of these bands (labeled as $\alpha$ y $\gamma$), are composed of Cu1 $d$, O1 $p$, 
Cu2 $d$ and O2 $p$ states. The other band (labeled as $\beta$) is composed of Cu2 $d$ and O2 $p$ states (see Table~\ref{Tabla3}). Note that the $\alpha$ and $\beta$ bands are 
nearly degenerate.

In the second region, in the X-$\overline{M}$ direction (see Fig.\ref{Fig.2a}(c)) there are three bands crossing the $E_F$. Two of these bands (labeled as 
$\alpha^\prime$ and $\gamma^\prime$) are composed of Cu1 $d$, O1 $p$, Cu2 $d$, and O2 $p$ states. The other band (labeled as $\beta^\prime$) is composed of Cu2 $d$ and O2 $p$ states.

Again around the $\overline{M}$ point in the $\Gamma$-Z direction (see Fig.\ref{Fig.2a}(a)), it is observed that the bands derived from Bi-O states do not cross $E_F$ and
are displaced above $E_F$ $\sim 230 meV$. Therefore in this case (form II) the Bi-O planes are nonmetallic. Now, below $E_F$ the bands derived from Cu-O states reach $\sim -270 meV$.

Comparing the band structures from form II with the ONES obtained from form Opt, it is observed a rigid displacement of $\sim 0.6 eV$ of the Bi-O bands toward higher energies.
As a result of this, all states associated to the bismuth atoms are now above $E_F$. Also the general behavior of the Cu-O bands below $E_F$ show no significant changes.

As a general behavior of the three band structures presented it is observed that the bands derived mainly from the Cu-O planes which cross the $E_F$ in the Z-X-$\overline{\text{M}}$
directions show no important differences in its dispersion, although some changes in its composition can be found (see Table~\ref{Tabla2} and Table~\ref{Tabla3}). These three
band structures present strongly two dimensional character.

\subsection{Fermi surface}

\begin{figure*}[htb!]
\begin{center}
  \begin{tabular}{ccc}
   \includegraphics[width=0.3\textwidth]{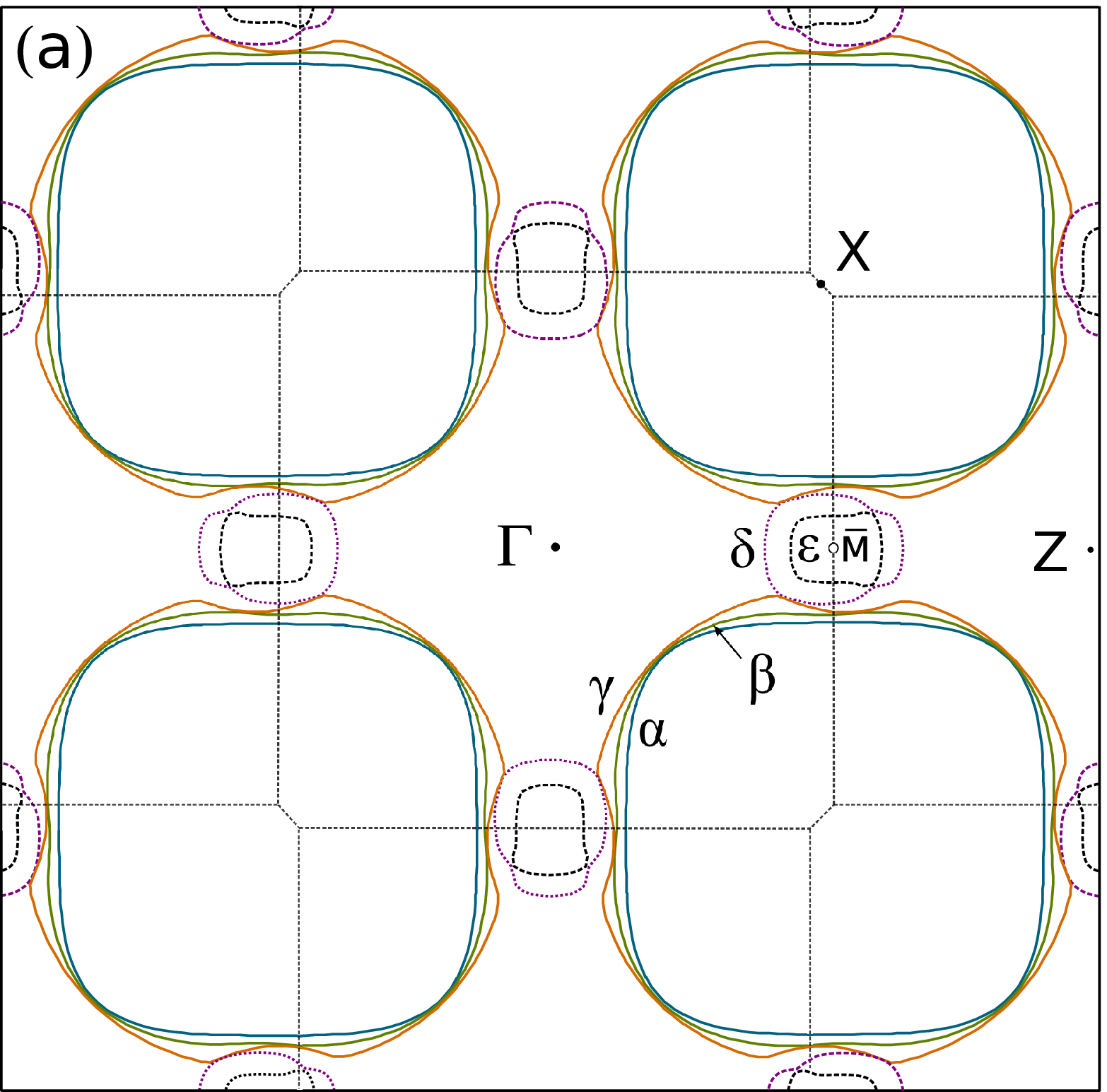} & \includegraphics[width=0.3\textwidth]{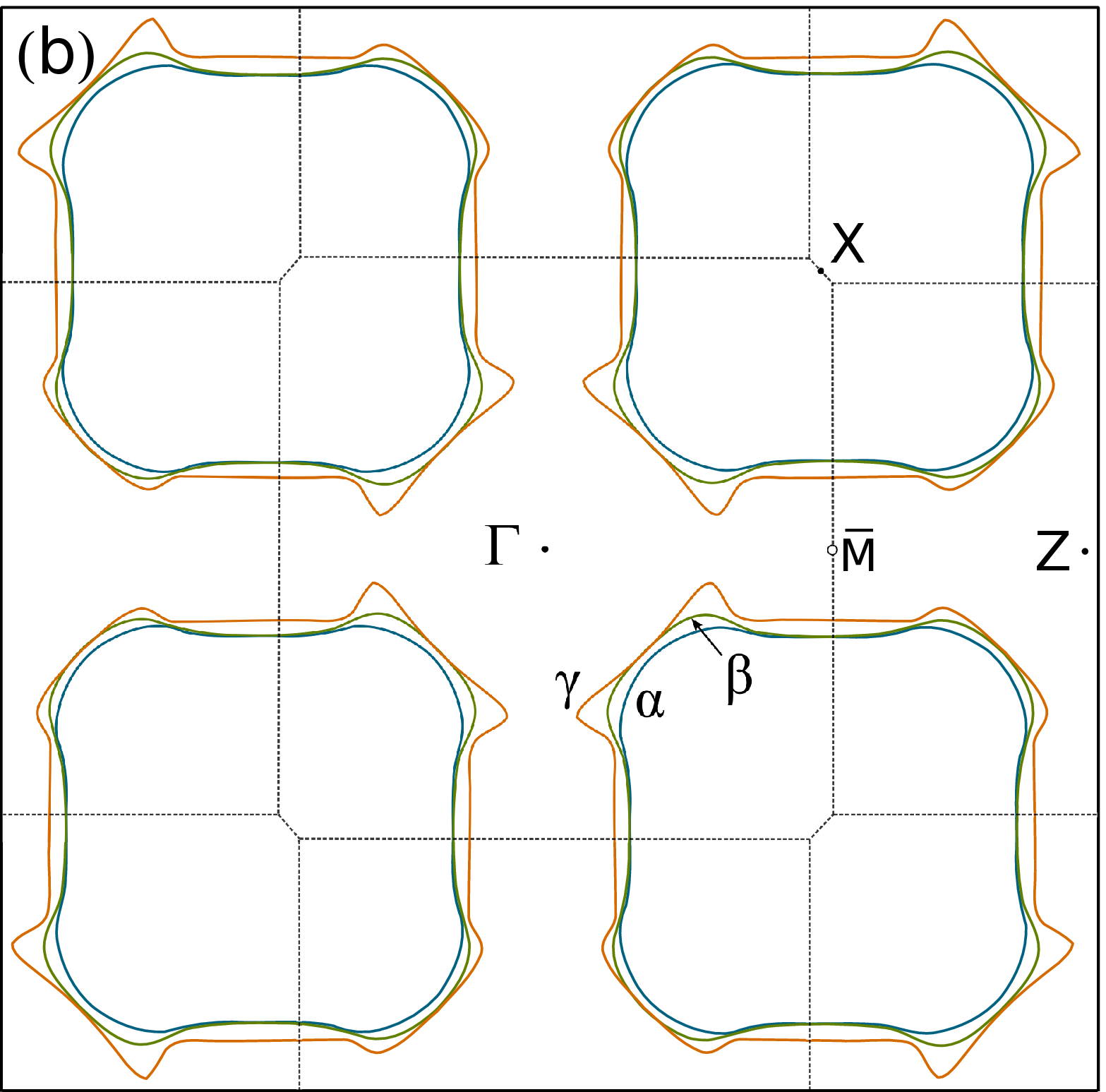} & \includegraphics[width=0.3\textwidth]{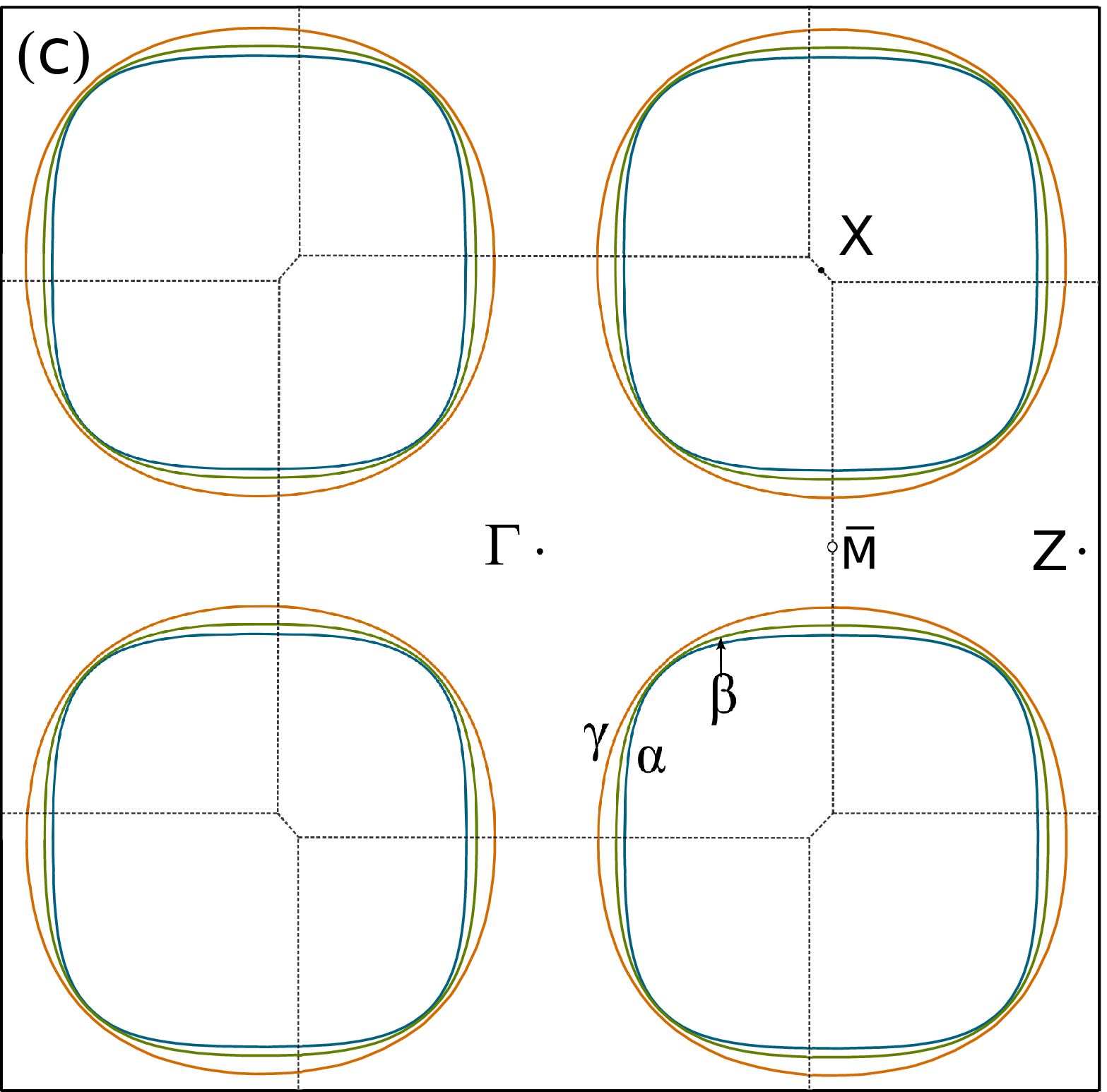} \\
\end{tabular}
 \end{center}
    \caption{\label{Fig.5} (Color online) The Fermi Surface (FS) at $k_z=0$ of $Bi_2Sr_2Ca_2Cu_3O_{10}$ in an extended zone scheme. (a) shows the FS for form Opt. The Bi-O pockets 
    are represented by violet and black lines (dashed lines). (b) AND (c) show the FS for form I and II respectively. These two FS's do not show Bi-O pockets.}
\end{figure*}

In Fig.\ref{Fig.5} we show the Fermi Surface (FS) of the Bi-2223 compound in an extended zone scheme for the three analyzed forms. A detailed study of the FS for form Opt is
presented in ref.~\cite{3a}. In this surface the important feature is the presence of the Bi-O pockets around the $\overline{\text{M}}$ point, that are in disagreement with 
the experimental data~\cite{4a,5a,6a,6aA}.

It is observed from Fig.\ref{Fig.5}(b) and Fig.\ref{Fig.5}(c) that the topology of the FS's is strongly affected by the atomic displacement of the O3. There are three not quite degenerate hole surfaces
around the X point, labeled $\alpha$, $\beta$, and $\gamma$. These surfaces are derived from Cu-O planes, although the form I has a contribution of the O3 atoms 
(see Table~\ref{Tabla2}). In these surfaces the most relevant feature is the complete absence of the Bi-O pockets around the $\overline{\text{M}}$ point, in agreement with the 
experimental results~\cite{4a,5a,6a,6aA}.

In Fig.\ref{Fig.5}(b) around the $\Gamma$ point spiky sheets appear in the $\gamma$ surface that are different from the one around the Z point, causing a lower two dimensional
character as compared to form II. Also in form I, in the X-$\overline{\text{M}}$ direction there is a presence of Bi-O states, contributing to the FS, which was observed in t
he band structure, as can be seen in Table~\ref{Tabla2}.

Mori et al.~\cite{13a} show that the number of bands crossing at $E_F$ (in the nodal direction) is proportional to the number of the Cu-O planes in the multilayer cuprates. LDA electronic calculations present the
same feature~\cite{3a}. The band structures from the forms I and II (see Fig.\ref{Fig.2}(b) and Fig.\ref{Fig.2a}(b)) show again these three bands crossing at $E_F$. Two of them are 
nearly degenerate. The three band splitting is not observed in the experiment due to the degeneration of these bands and the resolution limitations inherent 
to the experimental equipment~\cite{3a,5a}.

\begin{figure*}[htb!]
     \includegraphics[width=0.8\textwidth]{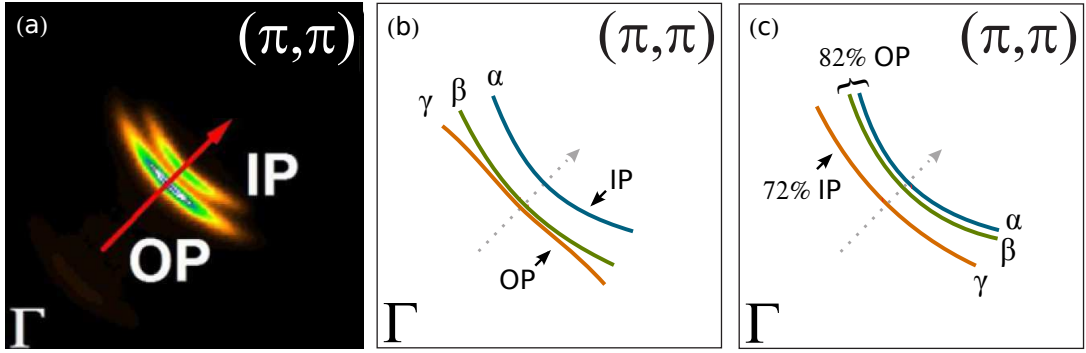} 
    \caption{\label{Fig.FS} (Color online) (a) The Fermi surface (FS) of Bi-2223 calculated by ARPES~\cite{5a} in the nodal direction. (b) and (c) Schematic FS showing the 
    band splitting in the forms I and II respectively.}
\end{figure*}

The FS of Bi-2223 calculated by angle-resolved photoemission spectroscopy (ARPES) is shown in ref.~\cite{5a}. In that work they found two surfaces on the $\Gamma$-X direction 
(see Fig.\ref{Fig.FS}(a)), that they assign to the outer copper planes Cu2-O2 (OP) and to the inner copper plane Cu1-O1 (IP) and suggest the possibility that the OP's are 
degenerate. Other experimental work with the same technique does not report this band splitting~\cite{6a,6aA}. 

In Fig.\ref{Fig.FS} the comparison between the experimental FS~\cite{5a} and the calculations done in this work for forms I and II in the nodal direction are shown.  
a good agreement with the experiment can be observed here.

In the FS of form I (Fig.\ref{Fig.FS}(b)) it is observed that the $\beta$ and $\gamma$ surfaces are nearly degenerate and derived from the Cu2-O2 planes (OP) with a contribution 
of the O3 $p_{x,y}$ states (see Table~\ref{Tabla2}), while the $\alpha$ surface is derived from the Cu1-O1 plane (IP). This result shows a good agreement with the composition of the 
surfaces assigned by Ideta et al. in the experiment reported in ref.~\cite{5a}.

In the FS of form II (Fig.\ref{Fig.FS}(c)) it is observed that the $\alpha$ and $\gamma$ surfaces are derived from the hybridization of the Cu2-O2 and Cu1-O1 planes (OP+IP), 
while the $\beta$ surfaces is derived from the Cu2-O2 planes (OP). In this case the $\alpha$ and $\beta$ surfaces are nearly degenerate, which are mainly derived from Cu2-O2 
planes (82\% OP), while the $\gamma$ surface is mainly derived from Cu1-O1 plane (72\% IP). The composition that we got from our calculations differs slightly from the ones assigned
by Ideta et al. 

Notice that form I agrees with the experimental assignment of the OP's as being degenerate(Fig.\ref{Fig.FS}(b)) . Form II (Fig.\ref{Fig.FS}(c)) indicates that it is the OP degenerate
surfaces, in disagreement with assignment of Ideta et al.~\cite{5a}.

Furthermore, nuclear magnetic resonance (NMR) studies of the Bi-2223 compound show that the hole concentration of the OP is found to be larger than that of the IP~\cite{13aA,13aB}, in 
agreement with our calculations in both cases.

\subsection{Density of states}
\begin{figure*}[hbt!]
\includegraphics[width=1\textwidth]{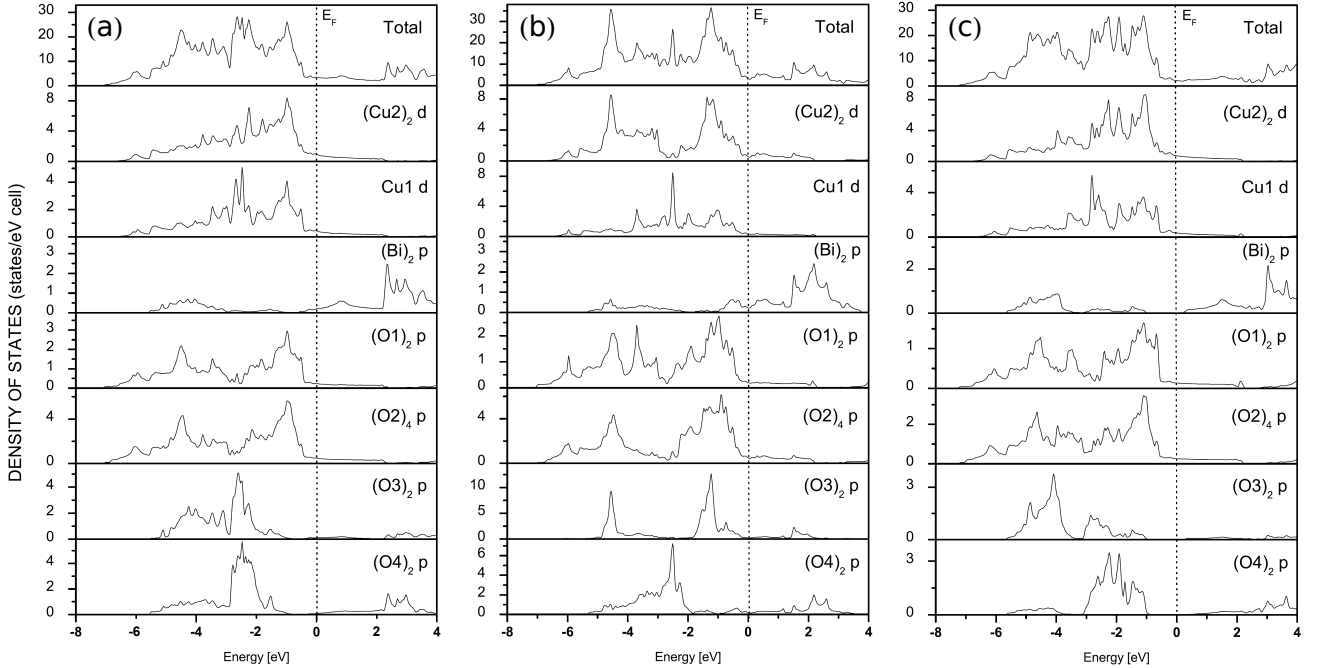} 
\caption{Total and atom-projected density of states for $Bi_2Sr_2Ca_2Cu_3O_{10}$ for the form Opt (a), form I (b) and, form II (c). Note the change of scale for each atom 
contribution.}
\label{Fig.1}
\end{figure*}

Fig.~\ref{Fig.1} shows both the total Density of States (DOS) and the atom-projected density of states (pDOS) for the forms Opt, I, and II. The analysis for the DOS of the form 
Opt is presented in the ref.~\cite{3a}.

As we just saw the displacement of the O3 atomic position in the crystalline structure generates significant changes in the band structure and in the FS. As expected, this changes 
are also visible in the DOS. Comparing the DOS of the form Opt with the form II we found that the general behavior is similar (as observed in the band structure and the FS).
On the contrary, important changes are observed in the DOS in the form I.

The most important feature in the DOS FOR the form II is the absence of any contribution of the Bi $p$, O3 $p$, and O4 $p$ states at $E_F$, a fact that leads to a lower density 
of states at $E_F$ (N($E_F$)) which is 2.15 states/(eV-cell) as compared to 3.55 states/(eV-cell) in form Opt and 3.14 states/(eV-cell) in form I. In Table~\ref{tabla1a} we 
presented both the atomic and the total contributions to the N($E_F$) for forms Opt, I, and II.

\begin{table*}[hbt!]
\caption{\label{tabla1a} Atomic contributions to the density of states at the $E_F$ ( N($E_F$)) for forms Opt, I, and II. The values are given in units of states/eV-atom. 
The total N($E_F$) is in units of states/eV-cell. The data for form Opt were taken from Ref.~\cite{3a}.}
\begin{tabular*}{0.6\textwidth}{@{\extracolsep{\fill}}lcccccccc}\hline\hline
& \multicolumn{7}{c}{Atomic state}&  Total  \\
\cline{2-8}
Compound & Cu2 $d$ & Cu1 $d$& Bi $p$ & O1 $p$ & O2 $p$& O3 $p$ & O4 $p$& N($E_F$)\\
\cline{1-1} \cline{2-2} \cline{3-3} \cline{4-4} \cline{5-5} \cline{6-6} \cline{7-7} \cline{8-8} \cline{9-9}
Form Opt  & 0.47    & 0.44   & 0.10   & 0.12   & 0.12  & 0.06   & 0.04& 3.55  \\
Form I    & 0.27    & 0.34   & 0.12   & 0.10   & 0.12  & 0.07   & 0.09& 3.14  \\
Form II   & 0.37    & 0.38   & -      & 0.07   & 0.06  & -      & -   & 2.15  \\
\hline\hline
\end{tabular*}
\end{table*}

In the DOS of the form I (see Fig.~\ref{Fig.1}(b)) it is observed that the Bi-O planes have a contribution at $E_F$, and as a consequence the Bi-O planes have a metallic character (see Table~\ref{tabla1a}). 
In the DOS of form II a shift with respect to the form Opt of the DOSp from Bi $p$ and O4 $p$ states towards higher energies from the $E_F$ is observed (see Fig.~\ref{Fig.1}(c)).
This results in a nonmetallic character of the Bi-O planes (see table~\ref{tabla1a}). 

An explanation for the metallic or nonmetallic character of the Bi-O planes can be understood as follows. The ionic character of the Bi atoms in the crystalline structure
tend to attract electrons into the Bi-O planes competing with the affinity for the electrons towards the Cu-O planes~\cite{7a}. This may involve the charge transfer between 
the Cu2-O2 and Bi-O4 planes. This charge transfer is possible by the interaction of the O3 atoms with the aforementioned planes. In the form I, when O3 is closer to the Cu2-O2 
planes it acts like a bridge for the charge transfer to the Bi-O4 planes, giving these planes the metallic character observed in one of our calculations. This explains the 
possible reason why the form I has a larger contribution of Bi-O states at $E_F$ than the form Opt (see Table~\ref{tabla1a}). On the other hand in the form II, O3 is closer to 
the Bi-O4 planes (away from the Cu2-O2 plane) which nullifies the charge transfer between these planes, explaining the nonmetallic character of the Bi-O planes.

We found two configurations that remove the pockets around the $\overline{M}$ point associated to the Bi-O planes. The relevant difference between these configurations 
lies in the metallic (form I) or nonmetallic (form II) character of the aforementioned planes.
 
Experimental results using scanning tunneling microscopy (STM), show that the Bi-O planes are nonmetallic in Bi-2212~\cite{14a,15a,16a}. 
To the best of our knowledge there are no experiments that define the metallic or nonmetallic character of the Bi-O planes in Bi-2223, except for the report presented by K. Asokan et al.
using X-ray absorption near edge structure (XANES) that shows a metallic character of the Bi-O planes for Bi-2223 and Bi-2212~\cite{17a}, the last one is in disagreement with the 
previous works just mentioned. However, the nature of the Bi-O planes character for Bi-2223 needs to be tested by more experiments.
The lack of enough experimental data for Bi-2223 does not allow us to choose which of the two studied configurations agrees completely with the experiments. Knowing the Bi-O planes 
character would permit us to support (or not) the assignation done by Ideta et al. in the FS compositions in the nodal direction~\cite{5a}. If the metallic case (Form I)
is supported by the experimental results, it would confirm that the surfaces derived from OP's are closer to the $\Gamma$ point which would agree with Ideta et al., while if 
the nonmetallic case happens, our calculations suggest that the surface that is closer to the $\Gamma$ point derives mainly from states associated to IP.  

Finally, we also calculated the total magnetic moment per cell for the forms Opt, I, and II and obtained $\sim$0.01 $\mu_B$, $\sim$0.02 $\mu_B$, $\sim$0.03 $\mu_B$ respectively. This 
implies that the compound does not exhibit a significant magnetic character at $T=0K$.

\section{Conclusions}

In this paper we addressed the so called Bi-O pockets problem, namely the appearance of states at the Fermi energy that are attributed to the Bi-O plane in theoretical ab initio 
calculations of the electronic band structure of Bi-cuprates and that are in contradiction with experiment. These pockets disappear in two ways either by using the experimental 
parameters and avoiding any optimization procedure (as we have checked) or by doping with 25\% of Pb~\cite{7a}. R. Kouba et al.~\cite{11bb} points to the need of the optimization procedure to get an 
electronic band structure that reproduces properly the electronic properties of the system in consideration. So, we have considered several ways in which the Bi-O pockets problem 
could be solved without doping and keeping the optimization procedure previous to the ab initio calculation. We performed small displacements of the positions of the Bi, O4, O3 
and Sr atoms in several configurations and found that some of them generate changes in the band structures that remove the Bi-O pockets that appear around the  
high-symmetry point $\overline{M}$. These pockets appear in all the theoretical ab initio electronic calculations of bismuth cuprates. The simplest way to remove the Bi-O pockets that we found is to displace 
the O3 atoms (the ones associated to the Sr-O planes) within two small intervals around its optimized position. Positions out of these small intervals lead to calculations that show 
the Bi-O pockets back. We studied the effect of an atomic displacement of the O3 atoms in two configurations which we called form I and form II in the electronic properties of 
$Bi_2Sr_2Ca_2Cu_3O_{10}$ using the LDA. In both cases the Bi-O pockets are removed. The relevant difference between these two configurations lies in the metallic (form I) or nonmetallic 
(form II) character of the Bi-O planes. This procedure is absent of any doping.
Our calculations for both forms present a good agreement with the experimental results calculated by angle-resolved photoemission spectroscopy (ARPES) and nuclear magnetic resonance 
(NMR). Since the experimental situation of the character (metallic or nonmetallic) of the Bi-O planes in Bi-2223 has not been totally clarified, we cannot decide which of the two forms 
reproduces more exactly the experimental results. Both, we emphasize, solve the Bi-O pockets problem.
A final question is of interest. Since the experiments determine the real position of every atom in the sample, the conclusion can be that either the position of the O3 atom is not 
define with enough precision yet by experiments or the slight displacement that corrects the results to describe the band structure in agreement with experiment, actually 
compensates an inherent error of the numerical methods of calculation or of the theory employed. It would be interesting to see if the appearance of the Bi-O pockets is inherent 
to all ab initio theories and methods of calculation (codes) existing nowadays.

\section{Acknowledgments}
The authors acknowledge to the GENERAL COORDINATION OF INFORMATION AND COMMUNICATIONS TECHNOLOGIES (CGSTIC) at CINVESTAV for providing HPC resources on the Hybrid Cluster 
Supercomputer ``Xiuhcoatl", that have contributed to the research results reported within this paper. J.A.C.M acknowledges the support of Conacyt M\'exico through a PhD scholarship.
D.E acknowledges the hospitality of the Department of Physics at Cinvestav.

\end{document}